# Improving the Understanding of Subsurface Structure and Dynamics of Solar Active Regions

**Principal Author:** S.C. Tripathy

National Solar Observatory, 3665 Discovery Dr., Boulder, CO 80303

**Coauthors:**

K. Jain[1], D. Braun[2], P. Cally[3], M. Dikpati[4], T. Felipe[5], R. Jain[6], S. Kholikov[1], E. Khomenko[4], R. Komm[1], J. Leibacher[1], V. Martinez-Pillet[1], A. Pevtsov[1], S. P. Rajaguru[7,9], M. Roth[8], H. Uitenbroek[1], J. Zhao[9]

1. National Solar Observatory, 3665 Discovery Drive, Boulder, CO 80303, USA
2. North West Research Associates, 3380 Mitchell Lane, Boulder, CO 80301, USA
3. School of Mathematics, Monash University, Clayton, Victoria 3800, Australia
4. High Altitude Observatory, NCAR, 3080 Center Green Dr., Boulder CO 80301, USA
5. Departamento de Astrofísica, Universidad de La Laguna, E-38205 La Laguna, Tenerife, Spain
6. School of Mathematics and Statistics, University of Sheffield, Sheffield S3 7RH, UK
7. Indian Institute of Astrophysics, Koramangala II Block, Bangalore 560034, India
8. Thüringer Landessternwarte, Sternwarte 5, 07778 Tautenburg, Germany
9. W.W. Hansen Experimental Physics Laboratory, Stanford University, Stanford, CA 94305, USA

**Category:   Basic Research (Solar Physics)**

## Synopsis

The goal of helioseismology is to provide accurate information about the Sun's interior from the observations of the wave field at its surface. In the last three decades, both global and local helioseismology studies have made significant advances and breakthroughs in solar physics. However 3-d mapping of the structure and dynamics of sunspots and active regions below the surface has been a challenging task and are among the long standing and intriguing puzzles in solar physics due to the complexity of the turbulent and dynamic nature of magnetized regions.

Thus the key problems that need to be addressed during the next decade are:

- Understanding the wave excitation mechanisms in the quiet Sun and magnetic regions
- Characterizing the wave propagation and transformation in strong and inclined magnetic field regions and understanding the magnetic portals in the chromosphere
- Improving helioseismology techniques and investigating the whole life cycle of active regions, from magnetic flux emergence to dissipation
- Detecting helioseismic signature of the magnetic flux of active regions before it becomes visible on the surface so as to provide warnings several days before the emergence

For a transformative progress on these problems require

- Full disk, simultaneous Doppler and vector magnetic field measurements of the photosphere up to the chromosphere with a spatial resolution of about 2 arc-sec
- Large-scale radiative MHD simulations of the plasma dynamics from the sub-photosphere to the chromosphere

Multi-height observations will also able to estimate and correct the center-to-limb variation which is crucial for the measurement of meridional flow in the deeper convection zone including the tachocline region. These measurements will further help to reduce the convective background noise in the power spectra resulting in a more accurate determination of the oscillation frequencies.

## 1. Introduction

Magnetic field at the Sun's surface is manifested in the forms of active regions, which are highly concentrated magnetic structures often surrounded by plage representing large areas of diffuse magnetic field. Although a deeper understanding of sunspots and active regions through the high-resolution observations and modeling has been a major area of investigation in solar physics, the mechanisms of their formation, evolution and eventual disintegration are still intriguing puzzles. It is well established that the solar dynamo responsible for creating the magnetic field and its rise through the convection zone are hidden from direct observations and can only be probed through helioseismology.

Recently, the understanding of the turbulent magnetized plasma in strong field regions has received a boost through radiative MHD numerical simulations and application of local helioseismology techniques to mid-to high-resolution observations. Forward modeling and numerical simulations suggest that active regions open a window from the interior into the solar atmosphere and that the seismic waves leak through this window. Under certain conditions the leaked acoustic waves are converted into additional MHD waves (Alfvén, fast/slow magnetic waves, etc.), reflect high in the atmosphere and re-enter the interior to rejoin the confined seismic wave field altering the original acoustic signal. Thus, further advancements in helioseismic inferences below the active region will require a precise understanding of interaction between the acoustic waves and the strong inclined magnetic field as a function of height in the solar atmosphere. Developing a complete understanding of this mode conversion would revolutionize the interpretation of helioseismic signals and open a new window into the properties of the magnetic field in active regions.

Therefore a better understanding of the complex processes associated with active regions can only be achieved with a synergy between forward modeling and helioseismic inferences from measurements through the application of local helioseismology. Recent theoretical and modeling studies have indicated that significant progress in this research area would require simultaneous vector magnetic field and high-cadence Doppler observations at multiple heights in the solar atmosphere to properly represent a transformation of acoustic waves to magneto-acoustic and Alfvén waves as the waves cross boundary between the convection zone and upper solar atmosphere. Such observations are also crucial to constrain theoretical models and to guide future developments in the theory.

Thus the fundamental questions about whether sunspots are the results of strong magnetic fields originating at the base of the convection zone, or they are formed in upper part of the convection zone via local instabilities, and how the instabilities of sunspot magnetic fields that lead to their decay cannot be fully answered without the application of observational input from multi-height helioseismology.

## 2. Scientific Motivation and Current Status

Both global and local helioseismic studies have significantly contributed to our knowledge of the solar interior. However, studies involving sunspots and active regions have not been well understood and are extremely challenging. The development of local helioseismic techniques of time-distance (TD), helioseismic holography (HH) and ring-diagram (RD) have provided measurements of travel times and oscillation frequencies associated with the subsurface structure and dynamics of sunspots. Despite the abundance of clues from observations at the solar surface and application of helioseismic techniques, theories about formation, subsurface structure, thermal properties and topology of active regions are still controversial. As demonstrated by Gizon et al. (2009), wave speed perturbations beneath an active region obtained from time-distance inversions are different from ring-diagram inversions and disagrees with the results from semi-empirical models or radiative MHD simulations. It is also suggested that the helioseismic inversions for sound speed beneath the sunspots are contaminated by surface effects associated with the sunspot magnetic field (Couvidat & Rajaguru 2007). In addition, power absorption in sunspots and enhanced acoustic power surrounding active regions, known as acoustic halos and acoustic glories are not yet fully understood. Below we highlight some of the recent progress made on this subject.

### 2.1 Forward Modeling

From a theoretical point of view the influence of magnetic fields on incident acoustic waves is a complex phenomenon and there is insufficient understanding of the processes involved. For example, the wave propagation through sunspots involves conversion between waves of acoustic and magnetic character (Cally 2007). Furthermore, rapid progress in numerical simulations of wave propagation in magnetized plasmas now provides better insight into the wave interaction with the magnetic field. Models show that the acoustic waves leak into the atmosphere through active regions and convert a significant amount of energy into the magnetic fast waves (Schunker et al. 2013). Under certain conditions, the fast waves are reflected back to the interior and in this process the phase of the acoustic waves (corresponding to travel times) is altered which leads to incorrect inferences. Similarly, scattering and mode mixing by the magnetic field causes modifications to the acoustic wave field (Hindman & Jain 2012) which manifest as a redistribution of power between different acoustic waves in the vicinity of a magnetized region. Thus, the separation of scattered and reflected waves from different heights is important for understanding the inhomogeneity in magnetic field regions above the surface.

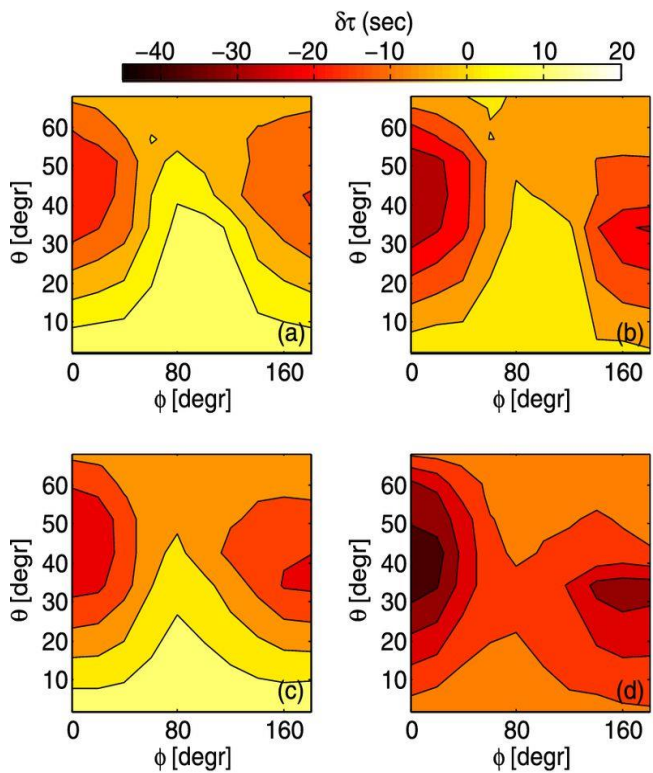

*Figure 1: These contour plots depict phase travel-time perturbations (δτ) of acoustic waves with respect to the quiet Sun model derived as function of wave source position (field inclination from vertical (θ)) and receiver location (azimuthal angle (φ)) for two different travel distances (from top to bottom 6.2, 8.7) for 3-mHz (left) and 5-mHz (right) frequency bands for a sunspot model with field strength of 1.5 KG. The figure demonstrates that the magnitude of δτ strongly depends on frequency and magnetic field inclination. Thus, the study advocates simultaneous Doppler and vector magnetic field observations at different heights to comprehend the interaction between the acoustic waves and inclined magnetic field. Adapted from Moradi et al. (2015).*

**2.2 Numerical Simulations**

The generation of artificial data through numerical modeling and its analysis through various local helioseismic techniques have also highlighted the issues associated with wave propagation in strong magnetized regions. Applying TD and HH methods to realistic MHD sunspot simulations, Braun at el. (2012) demonstrated that standard modeling efforts, assuming only a perturbation to wave speed, do not reproduce the expected travel time signatures. It is suggested that the inversion methods which incorporate direct effects of the magnetic field, including mode conversion (Crouch et al. 2011), are required to make further progress. The study of Cally & Moradi (2013) indicated that the propagating fast and Alfvén waves introduce a significant travel time shift that depends on the propagation direction of the waves and magnetic field inclination. The numerical simulation of Moradi et al. (2015) using model sunspots further confirmed this frequency-dependent directional behavior, consistent with the signature of magneto-hydrodynamic mode conversion (Figure 1) regardless of the sunspot field strength or depth of its Wilson depression.

Felipe et al. (2016, 2017) numerically simulated the propagation of a stochastic wave field through sunspot models with different Wilson depths and umbral magnetic field strength. The resulting simulations when analyzed through the HH techniques supported the frequency dependent directional behavior. The study also found that in some ranges of phase speed and frequency, the travel time shifts are primarily caused by changes in the path of the waves associated with the Wilson depression rather than the wave speed suggesting that inversions for the subsurface structure of sunspots must account for local changes in the density. All these investigations clearly point to the importance of understanding the wave interactions in the immediate vicinity of strong and inclined magnetic field especially at the height where the magnetic field dominates the plasma motion.

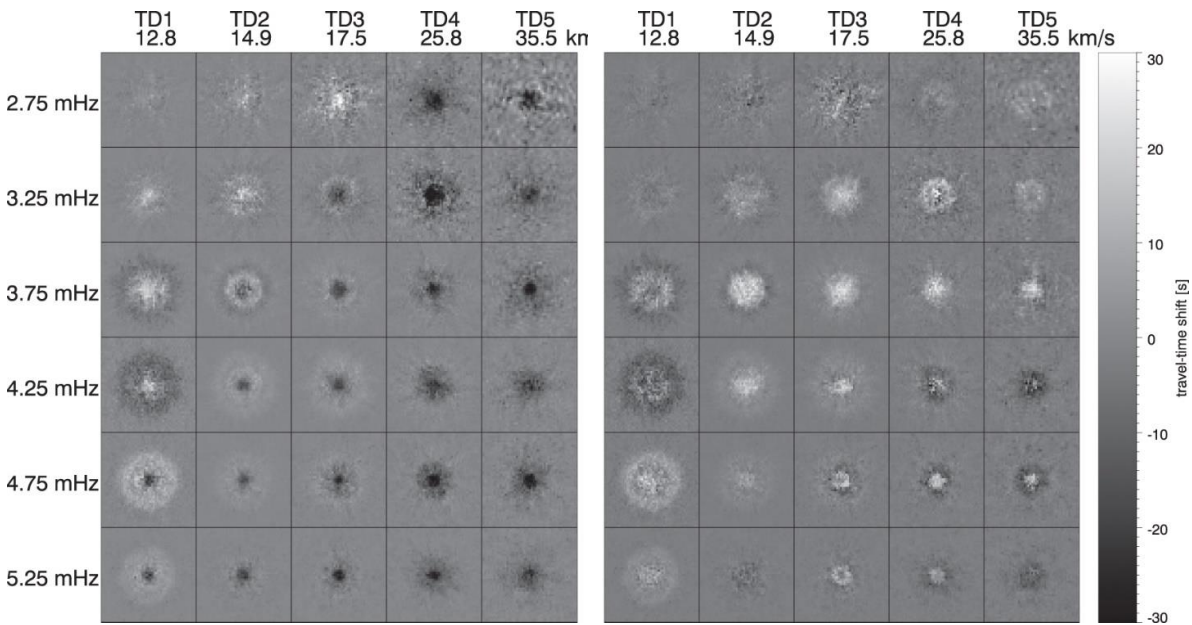

**Figure 2:** *Measured mean travel time shifts (with respect to a quiet region) for different frequency bands (y-axis) for five different phase-speed filters (shown in the top of the figures) using HH technique where wave propagation through sunspot models with different properties were numerically carried out. The left panel shows the measured travel time shifts for the full simulation (thermal plus magnetic) while the right panel shows the measured shifts only from magnetic simulation. In case of magnetic only sunspot model where the waves travel faster, it is expected that the travel time shifts should be mostly negative but found to be mostly positive. This emphasizes the importance for understanding the wave interactions in the vicinity of strong magnetic field regions (Adapted from Felipe et al. 2016).*

## 2.3 Acoustic Halos

The acoustic halos were first observed in Dopplergrams as a power enhancement around active regions at frequencies higher than the cutoff frequency (Brown 1992). Several studies since then have revealed new features but a successful physical mechanism to interpret these phenomena is yet to emerge. Khomenko & Collados (2009) suggested several potentially observable properties of halos, derived from their model of fast magneto-acoustic wave refraction, to be checked in future observations. Numerical simulations of power halos in MHS sunspot model suggest that the halo is solely produced by the return of the reflected fast magneto-acoustic waves (Rijs et al. 2016, also see Khomenko & Collados, 2009). This is based on a numerical experiment where the high-chromosphere Alfvén speed is artificially limited thereby preventing fast wave reflection, which is found to completely suppress halos, dramatically illustrating that the atmosphere can indeed shape the observed surface seismology. However, why these reflected fast waves are at a higher frequency than the cutoff frequency is still unclear. Also, the power enhancements above the cutoff frequency around the active regions are accompanied by power suppression below the cutoff frequency within the same active regions (Jain & Haber 2002) and the magnitude of such power suppression increases with height just above the surface (Jain et al. 2014). The variation of acoustic power in active regions may have significant effects on inferences of subsurface flows, because the suppression of acoustic sources in magnetized regions causes anisotropy in wave propagation properties. Since halos have been seen at many different heights of the solar atmosphere (Rajaguru et al. 2013; Tripathy et al. 2018), these questions again emphasize the need for multi-height observations.

## 2.4 Anomaly in Measured Phase shifts

In a recent study, Zhao et al. (2022) demonstrated that the phase shifts calculated between Doppler velocities at different atmospheric heights is a function of the wave frequency. For evanescent acoustic waves, oscillations in the higher atmosphere lead those in the lower atmosphere by about 1 second when their frequencies are below about 3 mHz and lags behind by about 1 second when their frequencies are above 3 mHz. Since it has been argued that the center-to-limb systematic effects arise due to the measurements at different atmospheric heights, it is postulated that the observed phase shifts, if not accounted for in time-distance measurements, will provide incorrect inferences. The study once again emphasizes the importance of multi-height Doppler observations to correctly measure the phase shifts which are interpreted as travel-time shifts.

## 2.5 Other Studies

Limited modern multi-spectral helioseismic observations with the MOTH instrument in Antarctica have been used to study p-mode characteristics in the presence of magnetic fields, and led to a discovery that the regions of inclined magnetic field can serve as "magnetic portals" through which acoustic energy can leak into the chromosphere (Jefferies et al. 2006). Multi-height helioseismic observations using HMI and AIA onboard SDO have shown the dependence of the wave power on the atmospheric height (Couvidat 2013, 2014). The analysis was restricted to intensity data (multi-height Doppler velocity data was not available) with limited resolution, but the results again stress the need for including multi-spectral observations in local helioseismic analyses. Using many different observables and photospheric vector magnetic field, Rajaguru et al. (2019) estimated the amount of wave energy that could be channeled from photosphere to chromosphere and concluded that much finer height coverage of the photosphere-chromosphere region is required to corroborate the role of small-scale magnetic fields in the wave heating of the solar chromosphere. An approach to obtain such a finer height coverage was already successfully carried out in recording Doppler velocity data on the quiet Sun (Wisniewska et al. 2016). This study pointed to the need to improve on the atmospheric models.

Therefore, any investigation advancing helioseismic inferences below the active region should include the physical effects of the magnetic field. The combination of analysis of observational data and numerical simulations will significantly enhance our understanding of the effect of the magnetic field on helioseismic signals and ultimately lead to more precise estimates of the subsurface structure and dynamics of the Sun in the presence of magnetic fields that are crucial to determine how and why the Sun varies. This may also enable forecasting the emergence of active regions to provide warnings several days in advance.

## 3. Critical Capabilities Needed for Future Helioseismic Studies of Sunspots and Active Regions

In summary, theory of linear propagation of the waves in strong and inclined magnetic field regions suggests that the incident fast acoustic wave from below the surface leaks into the higher atmosphere through the magnetized regions and under certain conditions these waves are converted into additional MHD waves (Alfvén, fast/slow magnetic waves, etc.). Numerical simulations of wave propagation in magnetized regions have further demonstrated the mode conversion process and have quantified the implications of the returning fast and Alfvén waves for the seismology of the photosphere. Thus, developing a complete understanding of this wave conversion and its effect on the helioseismic signals in the presence of active regions require contemporaneous observations at different heights in the solar atmosphere. The critical missing data, at present, are simultaneous vector magnetic field and high-cadence Doppler observations at multiple heights. Such observations are crucial for improving the understanding of acoustic wave propagation in the presence of magnetic field and decoding the structure and dynamics of sunspot and active regions in the subsurface layers. In addition, high spatial, spectral, and temporal resolution observations over small areas will also provide information on the unresolved small-scale processes that may be important.

The required observations can be obtained from both ground-based and space-borne observatories. It is worth mentioning that no space mission is being planned to fulfil these requirements. However, National Solar Observatory (NSO) is advancing the concept of a new global ground-based network for uninterrupted solar observations as there is strong interest in a modern solar synoptic network within the solar physics community both in the US and within the broader international solar community. In this context, NSO and the High Altitude Observatory have jointly prepared a proposal to design the next generation ground-based network (ngGONG), which will replace the existing Global Oscillation Network Group (GONG) and the Synoptic Optical Long-term Investigations of the Sun (SOLIS) instrument suite. Once operational ngGONG will provide key measurements of the solar atmosphere that drive the heliosphere and space weather as a single system (see White Paper by Pevtsov et al. (2022): ngGONG -- Future Ground-based Facilities for Research in Heliophysics and Space Weather Operational Forecast).

## 4. Observational Requirements

ngGONG with state-of-the art instrumentation could provide the necessary Doppler and vector magnetic field measurements at different heights in the solar atmosphere. Since the solar interior and magnetic field vary from cycle to cycle; advances in our understanding of active regions require long-term synoptic observations of the conditions inside the Sun. Thus, we propose that with additional financial support, ngGONG should be designed to measure Doppler velocity and vector magnetic field at different heights in the solar atmosphere (also see White Paper by Bertello et al. (2022): Multi-height measurements of the solar vector magnetic field). Since time-distance measurements are typically done with continuous observations over 8 hours for a single active region, it is necessary to carry out Doppler observations continually. However, the vector magnetic field observations could be carried out at a slower cadence. This will provide the statistics on the subsurface structure and dynamics of active regions that will be needed to evaluate the probability that a given active region will produce a geo-effective event. A recent study based on 18 years of GONG data clearly demonstrates that a high duty cycle required for such studies can be comfortably achieved with a network of ground-based observatories (Jain et al. 2021).

A similar science objective is considered by the European SOLARNET Solar Physics Research Integrated Network Group (SPRING) project led by the Leibniz-Institut für Sonnenphysik (formerly Kiepenheuer-Institut für Sonnenphysik) that developed a science requirement document and a conceptual design for the helioseismic Doppler Imager and magnetograph during 2015-2017. Currently, a preliminary design study for a network node carrying such a Doppler Imager is being pursued. In the context of ngGONG several meetings were also organized as tag on meetings to SHINE and other workshops in Boulder, USA.

Based on SPRING and discussions at the various tag on scientific meetings, the following observational requirements have been identified:

- Multi-height seismology proposed here could be achieved with a 2k × 2k detector. However, synoptic studies of active regions and magnetic fields would benefit from full-disk observations at 1 arc-sec spatial resolution (0.5 arc-sec/pixel) which would require 4k × 4k detectors. Advances in CMOS sensors appears to be a possibility. The higher resolution images would also require some type of image correction in order to achieve the full spatial resolution due to the Earth's atmosphere. One possibility is to use a tip-tilt mirror and high-speed computing to perform differential destretching of the data as it is collected.

- The necessary velocity and polarimetric sensitivity would require an entrance aperture of 20 to 40 cm, respectively.

- High temporal cadence of about 30 second in Doppler observations complementing the existing GONG and HMI spectral lines.

- Full disk vector magnetic field measurements at multiple heights.

- The choice of spectral lines should be such that their formation height samples the distinct heights in the solar atmosphere. There are a dozen spectral lines that meet the criterion including those used in GONG (Ni I 676.8 nm) and HMI (Fe I; 617.3 nm) observations. There are also several chromospheric line, potentially suitable for ngGONG instrument. One of the widely used lines is Ca II 854.2 nm line. Simplified inversion based on modified Milne-Eddington approximation was shown to be able to reproduce rather satisfactory chromospheric velocity and magnetic field extracted from Ca II 854.2 line and from Mg I b2 line (Dorantes-Monteagudo et al. 2022).